\documentclass[showpacs,preprintnumbers,preprint,nofootinbib]{revtex4}
\usepackage{amssymb}
\usepackage{amsmath}
\usepackage{graphicx}
\usepackage{dcolumn}
\usepackage{bm}
\usepackage{epsf}

\def\be{\begin{eqnarray}}
\def\ee{\end{eqnarray}}
\def\bea{\begin{eqnarray}}
\def\eea{\end{eqnarray}}

\def\H#1{{}^{#1}\mbox{H}}
\def\He#1{{}^{#1}\mbox{He}}

\begin{document}

\preprint{HEP/123-qed}
\title{Elastic proton scattering on tritium below the n-${}^{3}\mbox{He}$
threshold}
\author{Rimantas Lazauskas}
\email{rimantas.lazauskas@ires.in2p3.fr}
\affiliation{IPHC, IN2P3-CNRS/Universit\'e Louis Pasteur BP 28, F-67037 Strasbourg Cedex
2, France}
\date{\today }

\begin{abstract}
Elastic proton scattering on ${}^{3}\mbox{H}$ nucleus is studied inbetween p-%
${}^{3}\mbox{H}$ and n-${}^{3}\mbox{He}$ thresholds, in the energy region
where $\alpha$-particles first excited state is inbeded in the continuum.
For this aim Faddeev-Yakubovski equations are solved in configuration space,
fully considering effects due to the isospin breaking as well as rigorously
treating Coulomb interaction. Different realistic nuclear Hamiltonians are tested,
elucidating open problems in nuclear interaction description.
\end{abstract}

\maketitle

\renewcommand{\thefootnote}{\#\arabic{footnote}} \setcounter{footnote}{0}

\section{Introduction}

During the last decade remarkable progress has been achieved in computational few-nucleon physics. First of all Green Function Monte-Carlo~\cite{Pieper:2002ne} and No Core Shell model~\cite{Navratil:2000ww} techniques have
been developed and perfectioned, enabling to describe bound nuclei containing up to several nucleons. Lately these techniques have been also adopted to treat the simplest case of  the elastic nucleon-nucleus scattering~\cite{Nollett:2006su,Quaglioni:2008sm}. Great progress have been achieved also in treating
Coulomb interaction for three and four nucleon scattering~\cite{Kievsky:2000eb,Deltuva:2005cc}, opening the new
frontiers to test the nuclear interaction models and three-nucleon force~(3NF) in particular.

\begin{figure}[h!]
\begin{center}
\mbox{\epsfxsize=10.0cm\epsffile{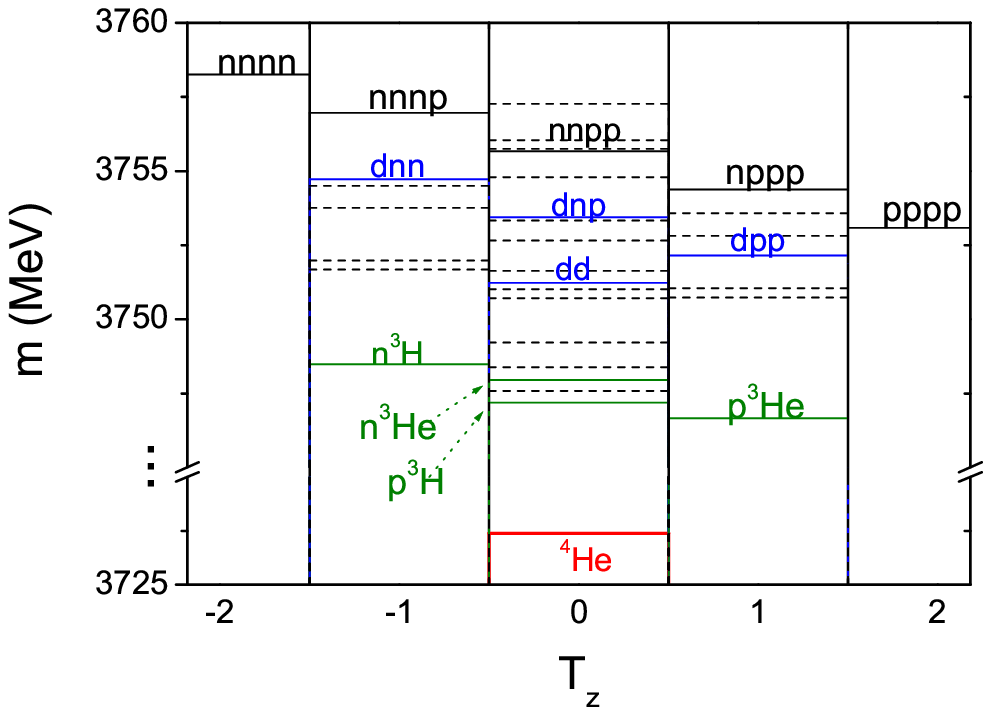}}\hspace{1.cm} %
\mbox{\epsfxsize=5.0cm\epsffile{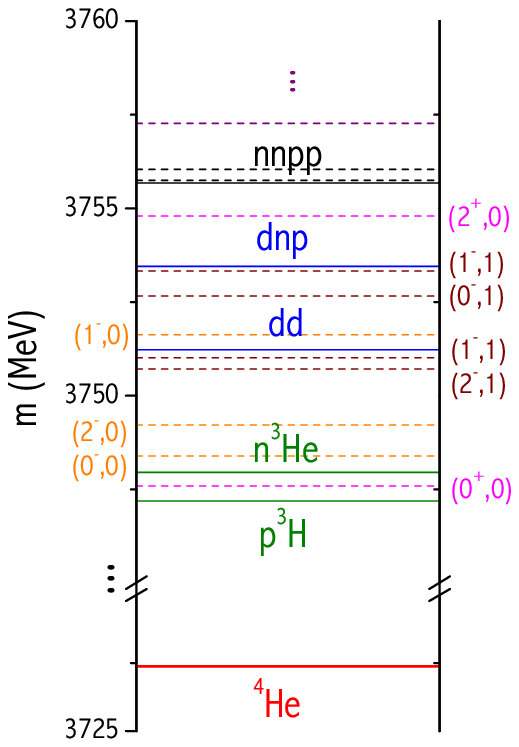}}
\end{center}
\caption{The experimental spectra of \textit{4N} bound and
resonant states are presented, as obtained using R-matrix analysis~\cite{Tilley_4N}.
In this figure resonances are indicated by the dashed lines. In the right pane of the figure
the structure of $\He4$ is elucidated. }
\label{Fig_4b_config}
\end{figure}

Regardless the continuous effort put to describe the strong nuclear interaction,
it still remains the central issue in nuclear physics. If the on
shell part of the nucleon-nucleon (NN) interaction is well constrained by the two
nucleon scattering data, determination of the off shell properties and
three-nucleon interaction in particular, which can be tested only in A$>$3
nuclei, is highly untrivial task. As a consequence much effort have been
devoted to describe the three nucleon system with special emphasis to N-d
scattering problem. However due to the large size of the deuteron, it
turns to be difficult to find observables, which reveal strong sensitivity to the off
shell structure of NN interaction. Low energy nucleon-deuteron scattering -- maybe apart
the J$^\pi$=1/2$^+$ state, where strong correlations with triton binding energy are
observed -- is well described by NN interaction alone and is quite insensitive to
its off shell structure. Only at large energies, well above deuteron
break-up threshold, off-shell effects become more pronounced~\cite{Witaa:2001by}. Unfortunately at
large energies one faces increasing difficulty in controlling the relativistic
effects and the large number of partial waves involved in the
scattering process~\cite{Witala:2004pv}. Finding relatively simple system at low energy, which could be
treated exactly and which is sensitive to the off-shell behavior of nuclear
interaction is of great interest. Four nucleon continuum containing series
of thresholds and resonances (see Fig.~\ref{Fig_4b_config}) seems to present an ideal laboratory.
Already the simplest case of n+${}^{3}\mbox{H}$ scattering, which is free of complications due to the Coulomb
interaction, contains several broad resonances in the continuum and presents
a serious test for nuclear interaction models: most of the realistic
Hamiltonians by $\sim10\%$ underpredict   elastic cross section
at the resonance region~\cite{Lazauskas:2004uq,Deltuva:2006sz,Deltuva:2008jr}.

The most interesting but also the most complex four nucleon structure is
continuum of the $^{4}$He, which contains numerous resonances and thresholds
(see Fig.~\ref{Fig_4b_config}). It has been suggested by Hofmann and Hale~\cite{Hofmann:2005iy} that
$\He4$ system can be used as a database to fine tune the three-nucleon interaction.
In particular delicate is the region between the p-$\H3$ and n-$\He3$ thresholds, due to the existence of
the $\mathcal{J}^{\pi }=0^{+}$ resonance in the p-$\H3$ threshold vicinity.
Accurate treatment of the charge symmetry breaking effects is required to separate
the two thresholds: as example if one neglects the Coulomb repulsion between the two protons
the resonance of the $\alpha$ particle moves below the p-$\H3$ threshold~\cite{These_Fred_97},
becoming a bound state and changing completely the nearthreshold scattering dynamics.
These facts reveal the particular relevance of this region in order to understand the isospin structure
of the two and three nucleon interactions.

This work aims to attract the attention to the p-$\H3$ elastic scattering problem, which
is studied for the proton laboratory energies up to
1 MeV, i.e. below n-$\He3$ threshold.  Six principally different realistic nuclear Hamiltonians will
be used to study this system.

\section{Faddeev-Yakubovski equations \label{sec_FY_eq}}

Four-nucleon problem is solved using Faddeev-Yakubovski (FY) equations in
configuration space~\cite{Fadd_art,Yakub}. In FY formalism four-particle
wave function is written as a sum of 18 amplitudes. From those we
distinguish amplitudes of type $K,$ which incorporate 3+1
 particle channels, while amplitudes of type H contain
asymptotes of 2+2 particle channels. By
interchanging order of the particles one can construct twelve different
amplitudes of the type $K$ and six amplitudes of the type $H$.

Further we use the isospin formalism, \textit{i.e.} we consider protons and
neutrons as being degenerate states of the same particle -- nucleon, having
the mass fixed to $\hbar ^{2}/m=41.47$ MeV$\cdot $fm$^{2}$. For the system
of the identical particles one has only two independent FYA, one of type $K$
and the other of type $H$. The other 16 FYA can be obtained by applying
particle permutation operators (\textit{i.e.} interchanging the order of
the particles in the system). Similarly only two independent FY equations exist;
by singling out $K\equiv K_{1,23}^{4}$ and $H\equiv H_{12}^{34}$ and
including three-nucleon force, FY equations read:~\cite{These_Rimas_03,Lazauskas:2008kz}:
\begin{eqnarray}
\left( E-H_{0}-V_{12}-\sum_{i<j}V_{ij}^{C}\right) K &=&V_{12}(P^{+}+P^{-})
\left[ (1+Q)K+H\right] +\frac{1}{2}[V_{23}^{1}+V_{31}^{2}]\Psi ,  \notag \\
\left( E-H_{0}-V_{12}-\sum_{i<j}V_{ij}^{C}\right) H &=&V_{12}\tilde{P}\left[
(1+Q)K+H\right] .  \label{FY1}
\end{eqnarray}%
Here $V_{12}$ is the strong part of the NN interaction between nucleons
(12), $V_{ij}^{C}$ is the Coulomb interaction between the nucleons $i$ and $j
$, the three-nucleon force (3NF) is represented by the terms $V_{23}^{1}$
and $V_{31}^{2}$, which are symmetric for the cyclic particle permutations
and which contain a part of 3NF acting in the particle cluster (123): $%
V_{123}$=$V_{12}^{3}+V_{23}^{1}+V_{31}^{2}$. The particle permutation
operators $P^{+}$, $P^{-}$, $\tilde{P}$ and $Q$ we use are simply:
\begin{eqnarray*}
P^{+}&=&(P^{-})^{-1}=P_{23}P_{12},\cr
Q&=&- P_{34},\cr
\tilde{P}&=&P_{13}P_{24}=P_{24}P_{13}.
\end{eqnarray*}
Employing operators defined above, the total wave function of the
four-nucleon is given by
\begin{equation}
\Psi =\left[ 1+(1+P^{+}+P^{-})Q\right] (1+P^{+}+P^{-})K+(1+P^{+}+P^{-})(1+%
\tilde{P})H.  \label{FY_wave_func}
\end{equation}

The numerical implementation of these equations is described in detail in~\cite{These_Rimas_03}.
This formalism enables to include Coulomb interaction as well as to test
different realistic nuclear interaction models, comprising non-local ones
and models in conjunction with three-nucleon interaction.

The $\He4$ system is predominantly total isospin $\mathcal{T}$=0 state.
However, the basis limited to $\mathcal{T}$=0 does not allow unambiguous separation
of the p-$\H3$ and n-$\He3$ channels. In order to separate these mirror
channels we allow for full isospin-symmetry breaking by incorporating the
total isospin $\mathcal{T}$=1 and $\mathcal{T}$=2 states in the partial wave basis.
This operation allows to account both for the effects due to the charge independence
and charge symmetry breaking in the strong part of the nuclear interaction.

\section{Results}

In this study proton scattering on $\H3$ nuclei for incident~(laboratory) proton energies up to E$_{p}=$1 MeV,
that is below n-$\He3$ threshold, is considered. This energy region is very delicate due to the
presence of the first excitation of the $\alpha$ particle, physically
situated, just above p-$\H3$ threshold, see Fig.~\ref{Fig_4b_config}. Subthreshold scattering cross section
is very sensitive to the precise position of this resonance.
Actually, the width of the resonance is strongly correlated with its
position relative to the p-$\H3$ threshold. If this state is slightly
overbound the resonance peak in excitation curve is naturally shifted to lower energies
and becomes more narrow. In case of underbinding one will have much broader resonance,
reflected in the flat excitation curve $\frac{d\sigma}{d\Omega}(E, \theta)$ as a function
of energy.
Very accurate description of the resonant scattering is therefore required.
In particular, special care should be taken of charge symmetry breaking terms in
order to properly separate the p-$\H3$ and n-$\He3$ thresholds.
Our scattering calculations, which do not restrain the total isospin $\mathcal{T}$,
fully accounts for the isospin symmetry breaking and thus are perfectly suited.

Several very different realistic nuclear Hamiltonians have been tested in this work.
Those included local configuration space potential of Argonne group AV18~\cite{POT_AV18}, non-local
configuration space potentials INOY~\cite{POT_INOY} and ISUJ~\cite{POT_ISUJ}, as well as chiral effective field theory based
potential of Idaho group derived up to next-to-next-to-next-to-leading order (I-N$^3$LO)~\cite{POT_IN3LO}.
Urbana three-nucleon interaction was also used in conjunction with AV18 and I-N$^3$LO potentials.
The parametrization, commonly called UIX~\cite{POT_UIX}, of this three-nucleon interaction was used together
with AV18 model for NN interaction; on the other hand the two-pion exchange parameter has been
assigned slightly different value A$_{2\pi}$=-0.03827 MeV  when using this force in conjunction with I-N$^3$LO NN potential.
The last adjustment allowed to fit the triton binding energy to its experimental value in a model we
refer I-N$^3$LO+UIX* in that follows.

Partial wave expansion is one of the most common ingredients in few-nucleon calculations. This expansion
converges rather fast for a low energy system, since the effective centrifugal terms grow
rapidly with the angular momentum. Nevertheless, due to
multitude of the degrees of freedom it represents, the partial wave basis for four nucleon system becomes considerable
and requires thousands of amplitudes to achieve numerically converged results. Therefore partial wave convergence
turns to be a central issue for
any calculation claiming numerical accuracy. In this study numerical convergence was studied as a
function of $j_{max}=max(j_x,j_y,j_z)$ for amplitudes K and $j_{max}=max(j_x,j_y,l_z)$ for amplitudes H,
where $j_x$, $j_y$, $j_z$ and $l_z$ are the partial angular momenta  as explained in~\cite{Lazauskas:2004uq}.
In Table~\ref{Convergence} convergence for p-$\H3$ scattering length calculations with INOY potential are presented. As one can
see convergence pattern is not regular, however final results seem to be converged better than at 0.5\%
level.
\begin{table}[htbp]
\caption{\label{Convergence} Convergence of p-$\H3$ scattering length calculations for INOY potential.
Scattering lengths are presented as a function of the maximal value of the partial angular momenta~($j_{max}$)
allowed in the calculations.}
\begin{ruledtabular}
\begin{tabular}{llll}
  $j_{max}$& $B(\H3)$&  $a_0$ & $a_1$ \\  \hline
    1   &     8.049  &  1.070 & 5.051 \\
    2   &     8.402  & -30.65 & 5.583 \\
    3   &     8.481  & -37.01 & 5.452 \\
    4   &     8.482  & -36.87 & 5.377 \\
	5   &     8.483  & -37.16 & 5.371 \\
    6   &     8.483  & -37.35 &       \\
\end{tabular}
\end{ruledtabular}
\end{table}

In the first three columns of the Table~\ref{Scat_lenght} calculated ground state binding energies for
three-nucleons and alpha particle are given. These values perfectly agree with
the ones obtained using other ab-initio methods.
The obtained results for triplet and singlet p-$\H3$ scattering lengths are also provided in
Table~\ref{Scat_lenght}. These values are estimated with some small error, which
is due to the fact that numerical calculations have been performed
for incident proton energies E$_p>7$ keV and then extrapolated to get corresponding
scattering lengths at E$_p=0$.
\begin{table}[htbp]
\caption{\label{Scat_lenght} Different nuclear model predictions for bound state energies of triton, $\He3$ and $\He4$
 in MeV together with p-$\H3$ scattering lengths in fm.}
\begin{ruledtabular}
\begin{tabular}{llllll}

  Model& B($\H3$) & B($\He3$) & B($\He4$) & $a_0(p-\H3)$ & $a_1(p-\H3)$\\  \hline
  ISUJ      &8.482&    7.718  & 28.91& -35.5(2)& 5.39(1)\\
  INOY      &8.483&    7.720  & 29.08& -37.4(2)& 5.37(1)\\
  AV18      &7.623&    6.925  & 24.23&  -15.5(1)& 5.79(1)\\
  AV18+UIX  &8.483&    7.753  & 28.47&  -23.6(2)& 5.47(1)\\
  I-N$^3$LO &7.852&    7.159  & 25.36&  -19.6(2)& 5.85(1)\\
  I-N$^3$O+UIX*&8.482& 7.737  & 28.12&  -22.9(2)& 5.59(1)\\\hline
  Exp.         & 8.482 & 7.718 & 28.30 & & \\
\end{tabular}
\end{ruledtabular}
\end{table}
 One can see important spread of the model predicted values, in particular for
the resonant singlet scattering length. As already explained this value is determined
by the position of the $\alpha$ particle excitation relative to the $p-\H3$ threshold, which
are separated roughly by $\backsim\hbar^2/(2\mu a_0^2)<$100 keV (here $\mu$ is reduced mass of the p-$\H3$ system).
Therefore only a few keV variation of the resonance position relative to p-$\H3$ threshold can
shift singlet scattering length by as much as 1 fm.
Though it is difficult to establish correlation laws for attractive four nucleon system, one should
stress sensitivity of the scattering length to the prediction of the  ground state binding energy of the alpha particle. As one
can see the 170 keV reduction in alpha particle binding energy obtained when passing from INOY to ISUJ models, permits to
reduce the resonant scattering length by almost 2 fm. This is not surprising however, once one shifts alpha
particle ground state relative to p-$\H3$ threshold -- its excitation is also shifted in the same direction.
\begin{figure}[tbp]
\includegraphics{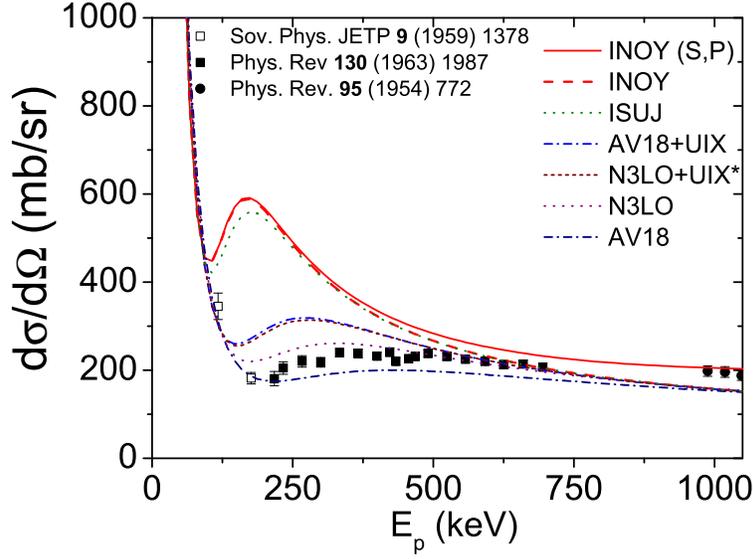}
\caption{\label{Fig:excrv}
Various model calculations for p-$\H3$ excitation function $\left.{\frac{%
d\sigma}{d\Omega}}(E)\right|_{\theta=120^{\circ}}$ compared with experimental data from~\cite{Jarmie,Ennis,Balashko}.}
\end{figure}

The p-$\H3$ triplet channel (J$^\pi=1^+$) is repulsive, while variation of the p-$\H3$ triplet scattering length is rather moderate:
this scattering length tends to decrease as three-nucleon binding energy increases.
Correlation is not perfect however, in particulary results for momentum space I-N$^3$LO potential tends to break the correlation.
Probably it is due to the fact that for this potential we have used point-Coulomb repulsion, while the configuration space
potentials employ screened Coulomb interaction having the same parametrization as
the one encoded in EM part of  the AV18 proton-proton potential.

\begin{figure}[h!]
\begin{center}
\mbox{\epsfxsize=7.50cm\epsffile{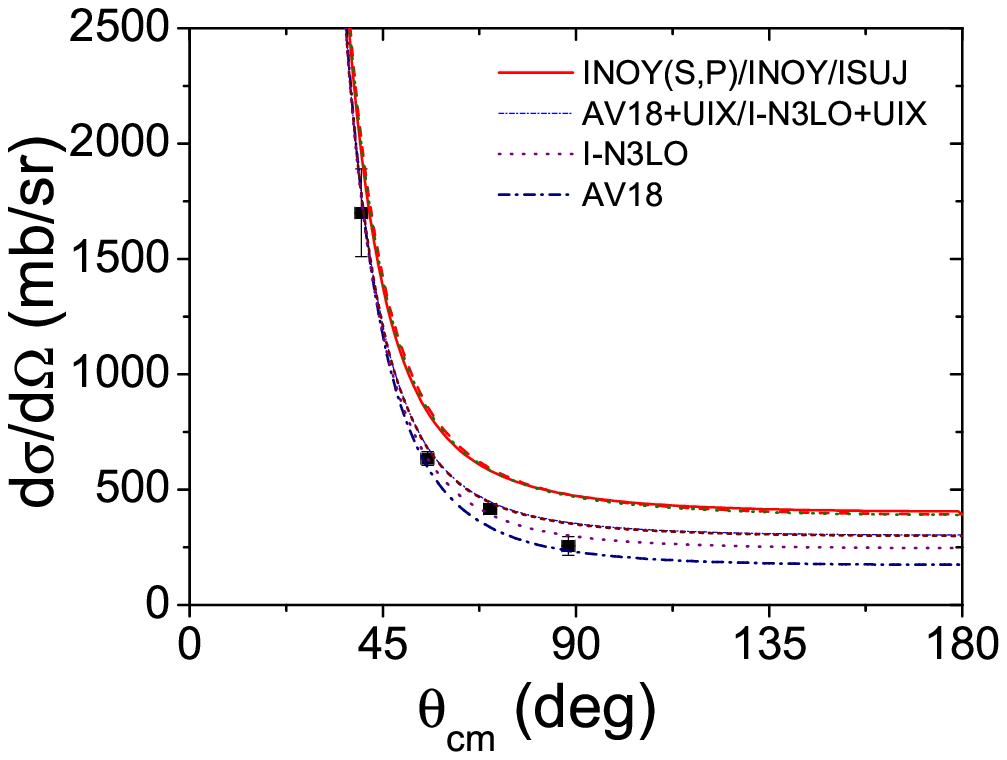}}\hspace{1.cm} %
\mbox{\epsfxsize=7.50cm\epsffile{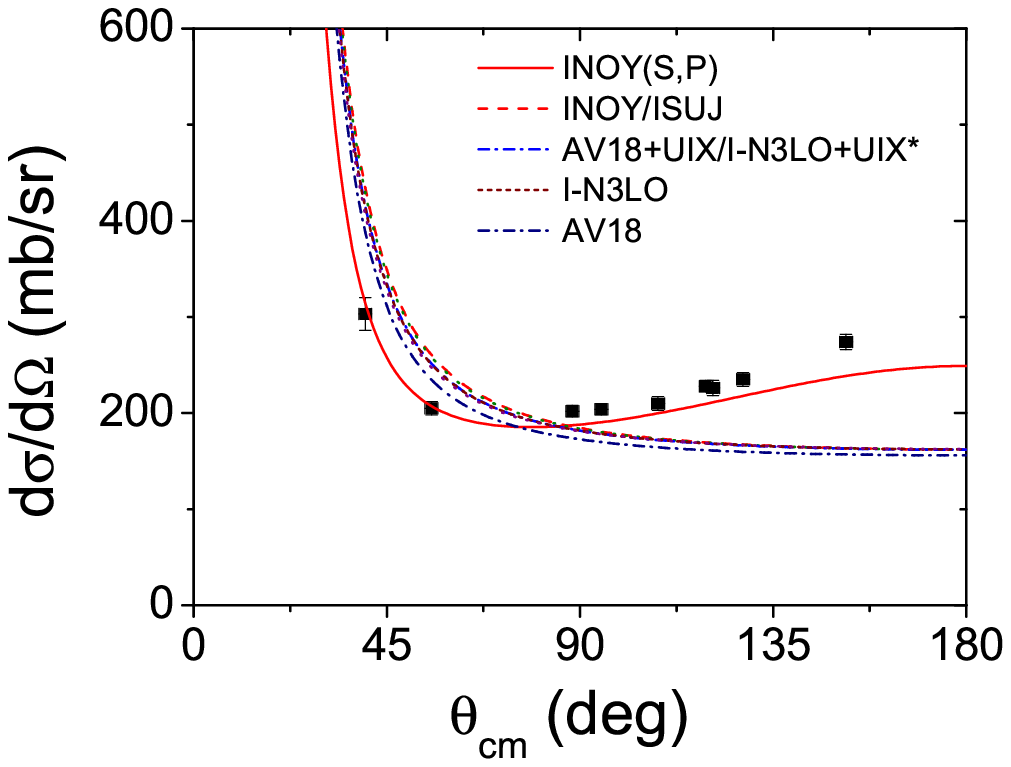}}
\end{center}
\caption{Calculated p-$\H3$ differential cross sections are compared with experimental data of~\cite{Balashko_df}
for incident proton energies of 0.4 MeV (left pane) and 0.9 MeV (right pane). If at 0.4 MeV contribution
of P-waves is negligible, they are necessary at 0.9 MeV in order to reproduce bending in the experimental
curve. On the other hand if S-wave contributions to the elastic scattering cross section differ from
model to model at 0.3 MeV, they merge the same curve at 0.9 MeV. }
\label{Fig:diffcs}
\end{figure}

In the Fig.~\ref{Fig:excrv} calculated p-$\H3$ differential cross section at $\theta_{cm}$=120$^\circ$ as a function of proton
laboratory energy is compared to the experimental data.
At very low energies, below 100 keV,  scattering cross section diverges due to the effective Coulomb repulsion between the proton
and triton nucleus. This behavior, which is easily described analytically, hides completely strong interaction effects.
Only beyond 100-150 keV nuclear scattering amplitudes are pronounced, revealing the clear resonant behavior of the singlet channel
(J$^\pi=0^+$), even though this
wave is suppressed by the statistical factor 3 compared to the triplet (J$^\pi=1^+$) one.
 The resonant peak is in particular neat for INOY and ISUJ models,
which places it too close to the  p-$\H3$ threshold together strongly underestimating its width.
The resonance width naturally increases as resonance moves further into continuum -- therefore other model predicted excitation
functions ($\frac{d\sigma}{d\Omega}(E, \theta=120^\circ)$) are much flatter than INOY or ISUJ ones.
Probably the best description of the experimental data in the resonance region ($E_p\sim$200-600 keV) is achieved by I-N3LO model.
The AV18 Hamiltonian slightly underpredicts experimental scattering length,
its excitation function is too flat and elastic scattering cross section is underestimated below 600 keV.
Inclusion of Urbana type three nucleon force makes I-N3LO+UIX* and AV18+UIX model results almost indiscernible, while
the elastic cross section is visibly overestimated in the resonance region. The agreement between two model predictions is
probably coincidental, as Hale and Hofmann\cite{Hofmann:2005iy} demonstrated strong sensitivity of the p-$\H3$ cross section to the 3NF
parameters. Qualitatively
the same behavior is observed, when analyzing the differential cross section as a function of the scattering angle.
It is demonstrated in the left pane of the figure~\ref{Fig:diffcs} for the incident protons of 300 keV.

Beyond the resonance region the S-wave cross sections for all models coincide. The AV18 prediction integrates into the joint curve the last, only for incident proton energies around 1 MeV. This is due to the fact that AV18 places the resonance too far into continuum, overestimating its width and resulting
extended in energy resonance region. The $\He4$ spectra contains several P-wave resonances just above n-$\He3$ threshold, see Fig.~\ref{Fig_4b_config}.
These resonances, and in particular narrow J$^\pi=0^-$ one, extend also into the region below n-$\He3$ threshold. Therefore
description of the scattering cross section  already beyond $E_p\approx$400 keV requires inclusion of the negative parity states.
This makes calculations largely involved, thats why calculations including P-waves have been performed only for INOY potential.
These waves seems to improve agreement with experimental data measured close to n-$\He3$ threshold.
This region is best to be studied analyzing differential cross section as a function of the scattering angle:
it is done in the right pane of the figure~\ref{Fig:diffcs} for incident protons of 900 keV,
which is only $\sim$100 keV below n-$\He3$ threshold. One can see that at this energy  S-wave contribution to the
differential cross section is almost the same for all the models considered. As demonstrated for INOY model:
P-waves, having in particular strong
contribution from the J$^\pi=0^-$ channel, are required in order to reproduce the bending of the experimental
cross section curve at the backward angles. Forward scattering cross section is however slightly underestimated by this model,
indicating that negative parity resonances are slightly displaced to higher energies by this interaction.
The similar observation have been made by Deltuva et al.~\cite{Deltuva:2007xw} for low energy n-$\He3$ scattering.
It is obvious that NN P-waves should have a strong impact for the negative parity states of the nuclear system and
therefore stronger NN P-waves than those given by the INOY model will be favorised by the experimental data.

Before closing this section possibility of using $\He4$ continuum as a laboratory to fine tune
the two and three nucleon interactions should be emphasized. Indeed, the p-$\H3$ scattering at very
low energies, having particular sensitivity to charge symmetry breaking terms in NN S-waves, 
is important to understand the charge symmetry breaking of the nuclear interaction.
At higher energies close to n-$\He3$ threshold contribution of negative parity states to the scattering process
can be well separated from the contribution of the positive parity ones. Negative parity states
turns to demonstrate stronger sensitivity to NN P-waves. Furthermore if the
study of the $\He4$ continuum is undertaken in parallel with resonant scattering in n-$\H3$ and p-$\He3$ systems
the new frontiers opens to understand isospin symmetry of NN P-waves as well as isospin structure of the
three-nucleon interaction. In view of such perspectives importance of low energy experiments in four nucleon
system should be stressed. In particular the nucleon scattering on tritium data is very scarce, while the
tritium experiments have been completely abandoned  for the last 30 years.
The nice exception is recently performed proton-tritium differential cross section
measurement in Fudan University (Shanghai)~\cite{China} for proton incident energies from 1.4 MeV to 3.4 MeV and
detectors fixed at the backward laboratory angle of 165$^\circ$.  Hopefully this activity is continued
for lower energy protons as well as extended to use the neutrons as projectile.

\section{Conclusion}
 In this work the proton elastic scattering on tritons is studied for incident proton energies below 1 MeV
 using Faddeev-Yakubovski equations in configuration space and by fully accounting for isospin
 breaking effects in nuclear interaction. Converged results are obtained for
 six qualitatively different realistic nuclear Hamiltonians. However none of the tested Hamiltonians have
 been able to reproduce the shape of the nearthreshold $J^\pi=0+$ resonance together with the
 ground state binding energy of the alpha particle.

 It has been demonstrated that $\He4$ continuum is an ideal laboratory to fine tune the two and three
 nucleon interaction models and
 the particular role this system can play in understanding the charge symmetry breaking.
 Recent advances in computational techniques allow very precise calculations for four-nucleon scattering problem
 at low energies,  therefore new four-nucleon scattering experiments and in particular ones concerning nucleon scattering
 on tritium are strongly anticipated.

\section*{Acknowledgement}

The numerical calculations have been performed at IDRIS (CNRS, France). We
thank the staff members of the IDRIS computer center for their constant
help.

\newpage
\bibliographystyle{plain}
\bibliography{apssamp}

\begin{thebibliography}{99}

\bibitem{Pieper:2002ne}
  S.~C.~Pieper, K.~Varga and R.~B.~Wiringa,
  Phys.\ Rev.\  C {\bf 66}, 044310 (2002)
  [arXiv:nucl-th/0206061].
\bibitem{Navratil:2000ww}
  P.~Navratil, J.~P.~Vary and B.~R.~Barrett,
  Phys.\ Rev.\ Lett.\  {\bf 84}, 5728 (2000)
  [arXiv:nucl-th/0004058].
\bibitem{Nollett:2006su}
  K.~M.~Nollett, S.~C.~Pieper, R.~B.~Wiringa, J.~Carlson and G.~M.~Hale,
  Phys.\ Rev.\ Lett.\  {\bf 99}, 022502 (2007)
  [arXiv:nucl-th/0612035].
\bibitem{Quaglioni:2008sm}
  S.~Quaglioni and P.~Navratil,
  Phys.\ Rev.\ Lett.\  {\bf 101}, 092501 (2008)
  [arXiv:0804.1560 [nucl-th]].
\bibitem{Kievsky:2000eb}
  A.~Kievsky, C.~R.~Brune and M.~Viviani,
  Phys.\ Lett.\  B {\bf 480}, 250 (2000)
  [arXiv:nucl-th/0003054].
\bibitem{Deltuva:2005cc}
  A.~Deltuva, A.~C.~Fonseca and P.~U.~Sauer,
  Phys.\ Rev.\  C {\bf 72}, 054004 (2005)
  [Erratum-ibid.\  C {\bf 72}, 059903 (2005)]
  [arXiv:nucl-th/0509034].
\bibitem{Witaa:2001by}
  H.~Witaa, W.~Gloeckle, J.~Golak, A.~Nogga, H.~Kamada, R.~Skibinski and J.~Kuros-Zonierczuk,
  Phys.\ Rev.\  C {\bf 63}, 024007 (2001).
\bibitem{Witala:2004pv}
  H.~Witala, J.~Golak, W.~Glockle and H.~Kamada,
  Phys.\ Rev.\  C {\bf 71}, 054001 (2005)
  [arXiv:nucl-th/0412063].
\bibitem{Lazauskas:2004uq}
  R.~Lazauskas, J.~Carbonell, A.~C.~Fonseca, M.~Viviani, A.~Kievsky and S.~Rosati,
  Phys.\ Rev.\  C {\bf 71}, 034004 (2005)
  [arXiv:nucl-th/0412089].
\bibitem{Deltuva:2008jr}
  A.~Deltuva, A.~C.~Fonseca and P.~U.~Sauer,
  Phys.\ Lett.\  B {\bf 660}, 471 (2008)
  [arXiv:0801.2743 [nucl-th]].
\bibitem{Deltuva:2006sz}
  A.~Deltuva and A.~C.~Fonseca,
  Phys.\ Rev.\  C {\bf 75}, 014005 (2007)
  [arXiv:nucl-th/0611029].
\bibitem{Hofmann:2005iy}
  H.~M.~Hofmann and G.~M.~Hale,
  Phys.\ Rev.\  C {\bf 77}, 044002 (2008)
  [arXiv:nucl-th/0512065].
\bibitem{Tilley_4N} D.R. Tilley, H.R. Weller and G.M. Hale, Nucl. Phys. A\textbf{541} (1992) 1.
\bibitem{These_Fred_97} F. Ciesielski, PhD thesis, Universit\'{e}  Joseph Fourier, Grenoble (1997).
\bibitem{Fadd_art} L.D. Faddeev: Zh. Eksp. Teor. Fiz. \textbf{39}, (1960)
(Wiley \&\ Sons Inc., 1972). 1459 [Sov. Phys. JETP \textbf{12}, (1961) 1014.
\bibitem{Yakub} O.A. Yakubowsky: Sov. J. Nucl. Phys. \textbf{5} (1967) 937.
\bibitem{These_Rimas_03} R. Lazauskas: PhD Thesis, Universit\'e Joseph
Fourier, Grenoble (2003); http://tel.ccsd.cnrs.fr/documents/archives0/00/00/41/78/
\bibitem{Lazauskas:2008kz}
  R.~Lazauskas,
  arXiv:0808.1650 [nucl-th]; Few-Body Syst, DOI 10.1007/s00601-008-0006-3.
\bibitem{POT_AV18} R.B. Wiringa, V.G.J. Stoks, R. Schiavilla, Phys. Rev.
\textbf{C51} (1995) 38.
\bibitem{POT_INOY} P. Doleschall, I. Borb\'ely, Z. Papp, W. Plessas, Phys.
Rev. \textbf{C67} (2003) 064005.
\bibitem{POT_ISUJ}
  P.~Doleschall,
  Phys.\ Rev.\  {\bf C77} (2008) 034002.
\bibitem{POT_IN3LO} D. R. Entem and R. Machleidt, Phys. Rev. \textbf{C68}
(2003) 041001(R).
\bibitem{POT_UIX} B.S. Pudliner, V.R. Pandharipande, J. Carlson and R.B.
Wiringa, Phys. Rev. Lett. \textbf{74} (1995) 4396.
\bibitem{Deltuva:2007xw}
  A.~Deltuva and A.~C.~Fonseca,
  Phys.\ Rev.\  C {\bf 76}, 021001 (2007)
  [arXiv:nucl-th/0703066].
\bibitem{Ennis} E.~M.~Ennis and A.~Hemmendinger, Phys. Rev. \textbf{95} (1954) 772.
\bibitem{Balashko} Y.~G.~Balashko et al., Sov. Phys. JETP \textbf{9} (1959) 1378.
\bibitem{Jarmie}N.~Jarmie et al., Phys. Rev. \textbf{130} (1963) 1987.
\bibitem{Balashko_df} Y.~G.~Balashko et al., Journ. Izv. Rossiiskoi Akademii Nauk, Ser. Fiz. \textbf{28} (1964) 1124;
Proc. Nuclear Physics Congress, Paris (1964) 255.
\bibitem{China} X.~J.~Xia, W.~Ding et al, J. Nucl.\ Instrum.\ Meth.\  B {\bf266} (2008) 705.

\end{thebibliography}

\end{document}